\documentclass[modern]{aastex631}

\usepackage{graphicx}
\usepackage{subfigure}
\usepackage{enumerate}
\usepackage{amsmath}
\usepackage{appendix}

\usepackage{mathtools}
\usepackage[warnings-off={mathtools-colon,mathtools-overbracket}]{unicode-math}

\usepackage{cleveref}

\usepackage{booktabs}
\usepackage{multirow}

\graphicspath{{figures/}}

\usepackage{xspace}

\newcommand{\dd}{\mathbf{d}}

\newcommand{\ICMARTpy}{\texttt{ICMARTpy}\xspace}

\begin{document}

\title{SSC Radiation in the ICMART Model: Spectral Simulations and Application to the Record-Breaking GRB 221009A}

\author[0000-0002-2694-3379]{Xueying Shao}
\affiliation{Institute for Frontier in Astronomy and Astrophysics, Beijing Normal University, Beijing 102206, People's Republic of China}
\affiliation{School of Physics and Astronomy, Beijing Normal University, Beijing 100875, People's Republic of China}

\author[0000-0002-3100-6558]{He Gao}
\affiliation{Institute for Frontier in Astronomy and Astrophysics, Beijing Normal University, Beijing 102206, People's Republic of China}
\affiliation{School of Physics and Astronomy, Beijing Normal University, Beijing 100875, People's Republic of China}

\correspondingauthor{He Gao}
\email{gaohe@bnu.edu.cn}

\begin{abstract}
    This paper presents simulations of the synchrotron self-Compton (SSC) spectrum within the Internal-Collision-induced Magnetic Reconnection and Turbulence (ICMART) model.
    We investigate how key parameters like the magnetization $\sigma_0$ shape the broadband spectral energy distribution by regulating the electron distribution and magnetic field strength.
    The overall spectrum typically comprises two components: synchrotron radiation peaking at $E_{\rm p}$ with a low-energy spectral index $\alpha$ between -1 and -1.5, and an SSC component peaking at $E_{\rm ssc}$.
    At high energies, Klein-Nishina suppression causes an exponential cutoff.
    The flux ratio Y between these components is critical: when Y is small, the SSC peak can be suppressed.
    Spectral features of the synchrotron component reveal the underlying physical conditions: harder spectra with $\alpha\sim-1$ indicate a large Y parameter and strong KN suppression.
    We find a positive correlation between Y and $\sigma_0$, contrasting with internal shock model predictions.
    Applied to GRB 221009A, our model suggests $\sigma_0\leq20$ can reproduce the MeV-TeV observations.
    This study underscores the value of combined MeV-TeV observations in probing GRB emission mechanisms.
\end{abstract}

\section{Introduction}
Gamma-ray bursts (GRBs) are the most luminous explosions in the universe.
Their bursty emission in the hard-X-ray/soft-$\gamma$-ray band is commonly referred as the “prompt emission” \citep{Zhang2018book}, which can be attributed to synchrotron radiation from relativistic electrons \citep{Meszaros1997a,Uhm2013,Toma2009,Cheng2020}.
Intriguingly, a subset of GRBs exhibit additional spectral components extending across multiple bands \citep{Guiriec2011,Guiriec2013,Guiriec2015}.
For instance, several GRBs, such as GRB 080319B and GRB 201223A, have been observed to exhibit correlated optical and $\gamma$-ray emission during the prompt phase \citep{Guiriec2016b}.
This behavior is commonly interpreted as the extension of synchrotron radiation originating from low-energy electrons or regions with strong magnetic fields \citep{Patricelli2011, Xin2023, Jin2023}.
Meanwhile, thanks to its rapid-response X-Ray Telescope \citep{XRT}, the Neil Gehrels Swift Observatory has captured early X-ray emission simultaneous with $\gamma$-rays in a number of GRBs \citep{Vaughan2006, Morris2007, Cusumano2007, Starling2008}.
The high degree of synchronization between their X-ray and $\gamma$-ray light curves suggests a common emission mechanism.
On the other hand, at GeV energies, several bursts including GRB 090510 and GRB 090902B have shown an additional rising power-law spectral component beyond 10 GeV.
Their broad-band spectra can be adequately modeled by the superposition of synchrotron and synchrotron self-Compton (SSC) radiation components \citep{Abdo2009a, Ackermann2010, Peter2012}.
A comprehensive analysis of the full broadband spectral energy distribution offers critical insights into the physical conditions of the emission region, such as the properties of the electron population and the magnetic field, which often remain degenerate in purely synchrotron-based models.

GRB 221009A, known as the Brightest Of All Time (BOAT), was detected at TeV energies by the Large High Altitude Air Shower Observatory (LHAASO), with more than 64,000 photons above 0.2 TeV collected within the first 3000 seconds \citep{LHAASO}.
This observation marks the first time the onset of the afterglow has been detected in the TeV band.
By treating this TeV emission as an upper limit on the flux during the prompt phase, \citet{Dai2023} analyzed the flux ratio between the synchrotron self-Compton (SSC) and synchrotron components within the internal shock model.
Their results indicate that a magnetization parameter $\sigma_0$ greater than 30 is required to satisfy the observational constraints.
Such a high $\sigma_0$ strongly supports a Poynting-flux-dominated jet, in which the prompt emission is primarily powered by magnetic dissipation processes.
Several mechanisms have been proposed to account for such energy dissipation, including MHD-condition-breakdown scenarios \citep{Usov1994, Zhang2002}, the radiation-dragged dissipation model \citep{Meszaros1997}, the slow-dissipation model \citep{Thompson1994, Drenkhahn2002, Giannios2008, Beniamini2017}, current-driven instabilities \citep{Lyutikov2003}, and forced magnetic reconnection \citep{Zhang2011, Mckinney2012, Lazarian2019}.

In this work, we focus on a representative model: the Internal-collision-induced Magnetic Reconnection and Turbulence (ICMART) model, first proposed by \citet{Zhang2011}.
This model posits that the central engine ejects multiple shells with $\sigma_0 > 1$, which exhibit variations in mass and velocity.
Mechanical collisions between these shells distort the magnetic field and trigger relativistic magnetohydrodynamic (MHD) turbulence in the interaction regions. This further twists the magnetic field lines, initiating a cascade of magnetic reconnection events.
Through reconnection, magnetic energy is efficiently converted into kinetic energy of particles, which then produce the observed prompt emission via synchrotron radiation and SSC scattering.

The ICMART model has been shown to effectively account for a wide range of observed prompt emission properties.
For example, it naturally explains the two-component variability observed in many GRB light curves \citep{Gao2013}:
the slow-varying envelope is attributed to the exponential growth in the number of mini-jets and the associated curvature effects, while the fast-varying pulses arise from variations in the luminosity, latitude, and directional angle of individual mini-jets during an ICMART event \citep{Zhang2014}.
A notable illustration of the model's explanatory power is provided by GRB 230307A, whose temporal and spectral characteristics---particularly an unusual dip in its light curve---are better reproduced by the ICMART model than by the internal shock model \citep{Yi2025}.
Moreover, \citet{Shao2022} demonstrated that synchrotron radiation within a decaying magnetic field can produce a Band-like spectrum with reasonable spectral indices.
This framework also naturally accounts for the observed spectral evolution: a “hard-to-soft” pattern of the peak energy $E_\text{p}$ results from the decay of the magnetic field due to bulk expansion, whereas an “intensity-tracking” pattern emerges from the superposition of multiple ICMART events \citep{Lu2012b, Shao2022}.
Furthermore, the model can reproduce key empirical relations in GRBs, such as the Yonetoku relation and the Liang relation \citep{Yonetoku2010, Liang2015}, under specific parameter conditions \citep{Shao2025}.

This work presents a comprehensive investigation of the high-energy radiation produced via SSC emission within the framework of the ICMART model.
Using the computational approach from section \ref{sec:method}, we employ the previously developed code \ICMARTpy \citep{Shao2022,Shao2025,Yi2025} to compute detailed SSC spectra and broadband emission.
We focus on characterizing the relationship between SSC and synchrotron spectral components and their evolution with key physical parameters.
As a central application, we simulate the spectrum of GRB 221009A within this framework and identify a set of parameters that closely match the observations.
The results are presented in section \ref{sec:result}, and further summarized and discussed in section \ref{sec:conclusion}.

\section{Method}
\label{sec:method}
In this work, we concentrate on calculating the synchrotron and SSC spectrums with \ICMARTpy while detailed discussion of the bulk dynamics is deferred to our previous works in \citet{Shao2022,Shao2025}.

We first introduce the physical image of the model briefly.
Within the ICMART framework, a succession of highly magnetized shells with similar magnetized factor but different mass, velocity and luminosity, is launched from the central engine.
Mechanical collisions between the later, faster shells and the earlier, slower ones would distort the magnetic field configurations and induce MHD turbulence.
The simulation of an ICMART event begins with a merged shell moving with mass $M_\text{bulk}$, Lorentz factor $\Gamma_0$, radius $R_0$ and magnetized factor $\sigma_0 \equiv E_\text{m,0}/E_\text{k,0}$, where $E_\text{m,0}$ and $E_\text{k,0}$ represent the initial magnetic and kinetic energy of the bulk, respectively.
During the event, the magnetic energy of the shell will be converted into particle energy through magnetic reconnections, with the ratio of the dissipated magnetic energy denoted as $f_\text{p}$.

As collisions proceed and the distortion of field configuration accumulate to a critical point, a magnetic reconnection will occur, which will continue to distort the magnetic field and trigger more reconnections.
Numerical simulations have demonstrated that the reconnection-driven magnetized turbulence could self-generate additional reconnections \citep{Takamoto2015,Kowal2017,Takamoto2018}, inferring that the number of magnetic reconnections would increase exponentially and constitute an ICMART event.
For each reconnection, the initial magnetic energy depleted by $(\sigma_0 - 1) / \sigma_0$, resulting in the magnetized factor drops from $\sigma_0$ to $1$ within the reconnection area.
The dissipated magnetic energy is used to eject a bipolar mini-jet with Lorentz factor $\gamma \approx \sqrt{1+\sigma_0}$ \citep{Zhang2014}.
In the simulation, three rest frames are considered: (1) the rest frame of the jet bulk, parameters in which are described with a single prime ($'$); (2) the rest frame of the mini-jet, parameters in which are described with double primes ($''$); (3) the rest frame of the observer, parameters in which are described with no prime.
For simplicity, we ignore the effects of cosmological expansion, treating the lab frame as stationary with respect to the observer frame.
Quantities in these three frames are related through two Doppler factors \citep{Shao2022}:
\begin{eqnarray}
    \mathcal{D}_1 &= [\Gamma(1-\beta_\text{bulk}\cos{\theta})]^{-1},   \\
    \mathcal{D}_2 &= [\gamma(1-\beta_\text{jet}\cos{\phi})]^{-1},
\end{eqnarray}
where $\Gamma$ is the Lorentz factor of the bulk, $\beta_\text{bulk}$ and $\beta_\text{jet}$ are the velocity parameters corresponding to $\Gamma$ and $\gamma$ respectively, $\theta$ is the angle between the line of sight and the direction of the bulk at the location of the mini-jet, and $\phi$ is the angle between the direction of the mini-jet and the direction of the bulk shell in the bulk comoving frame.

Within each magnetic reconnection, particles are accelerated and collide with the surrounding matter, causing part of the electrons to be re-accelerated through Fermi acceleration into a power-law distribution given by
\begin{equation}
    Q(\gamma_e) = Q_0(\frac{\gamma_e}{\gamma_\text{e,m}})^{-p}, \gamma_\text{e,m}<\gamma_e<\gamma_\text{e,M},
    \label{eq:electron number distribution}
\end{equation}
where $Q_0$ is a normalization factor and $\gamma_\text{e,m}$ is the minimum injected electron Lorentz factor and both of them can be determined by solving the electron number and energy conservation equation \citep{Shao2025}.
The maximum injected electron Lorentz factor is defined as $\gamma_\text{e,M} \equiv \sqrt{\frac{6\pi q_e}{\sigma_TB''_\text{e}}}$ where $B''_\text{e}$ is the magnetic field strength in the emission region within the jet frame, $q_e$ and $\sigma_T$ are the electron charge and Thomson scattering cross section respectively.
In the absence of numerical simulation for magnetic turbulence and reconnection, we introduce a free parameter $k$ to relate $B''_\text{e}$ to the bulk magnetic field strength $B'$ as $B''_\text{e, 0} = \sqrt{k}B'/\gamma$.
Both $B''_\text{e}$ and $B'$ decay as a power law with respect to the bulk radius, with exponents of $-b$ and $-1$, where $b$ is significantly greater than $1$ due to the substantial consumption of magnetic energy during the reconnection process.

After Fermi acceleration, the electrons undergo radiative and adiabatic cooling.
The synchrotron cooling and adiabatic cooling are calculated following \citet{Uhm2013}, while the SSC cooling is calculated through a numerical iteration procedure.
In the rest frame of the mini-jet the synchrotron radiation power could be expressed as \citet{Rybicki1985}
\begin{equation}
    \label{eq:pnup}
    P''_\text{syn}(\nu'') = \frac{\sqrt{3} q_e^3 B''_\text{e}}{m_{\rm e} c^2}
    \int_{\gamma_{\rm e,m}}^{\gamma_{\rm e,M}}
    \left( \frac{dN_{\rm e}''}{d\gamma_{\rm e}} \right)
    F\left(\frac{\nu ''}{\nu_{\rm cr}''} \right) d\gamma_{\rm e},
\end{equation}
where $m_{\rm e}$ and $c$ are the electron mass and the speed of light respectively, $\nu_{\rm cr}'' = 3 \gamma_{\rm e}^2 q_e B''_\text{e} / (4 \pi m_{\rm e} c)$ is the characteristic frequency of an electron with Lorentz factor $\gamma_e$, $N_{\rm e}''$ is the electron distribution after cooling and $F(\frac{\nu ''}{\nu_{\rm cr}''})$ is the synchrotron spectrum of specific frequency.

In addition to synchrotron radiation, SSC is also considered in this paper.
The radiation power of the SSC spectrum is calculated as \citet{Blumenthal1970}:
\begin{equation}
    P''_\text{ssc}(\nu'') = \int N_{\rm e}''(\gamma_{\rm e})\dd \gamma_{\rm e} \int \frac{3\sigma_T m_{\rm e} c^3}{4\gamma_{\rm e}} g(\nu''_{\rm syn}, \gamma_{\rm e}) \frac{n_{\rm ph}(\nu''_{\rm syn})\dd \nu''_{\rm syn}}{\nu''_{\rm syn}},
\end{equation}
where $n_{\rm ph}(\nu''_{\rm syn}) = P''_\text{syn}(\nu'') / (h\nu'')$ is the number density of photons with $h$ being the Plank constant.
$g(\nu''_{\rm syn}, \gamma_{\rm e})$ is the cross section for scattering considering Klein-Nishina (KN) effect and is calculated as \citet{Blumenthal1970}
\begin{equation}
    g(\nu''_{\rm syn}, \gamma_{\rm e}) = 2q\ln{q} + (1+2q)(1-q) + \frac12\frac{(\Gamma_\epsilon q)^2}{1 + \Gamma_\epsilon q}(1-q),
\end{equation}
where $q=E_1 / (\Gamma_\epsilon(1-E_1))$, $E_1 = h\nu''/(\gamma_{\rm e}m_{\rm e}c^2)$, $\Gamma_\epsilon = 4h\nu''_{\rm syn}\gamma_{\rm e}/(m_{\rm e}c^2)$.
The broadband spectrum is therefore the superposition of the synchrotron and SSC spectrums of all the magnetic reconnections:
\begin{equation}
    P_{\nu, \text{tot}}= \sum_{N_{\text{tot}}}\mathcal{D}_1^3 \mathcal{D}_2^3 (P''_\text{syn}(\nu'') + P''_\text{ssc}(\nu'')),
\end{equation}
where $N_{\text{tot}}$ is number of the magnetic reconnections.

\section{Result}
\label{sec:result}
\subsection{Spectral Shape}
\label{subsec:spectral shape}
\begin{figure}
    \centering
    \subfigure[General shape \label{fig:sample}]{
        \includegraphics[width=0.45\textwidth]{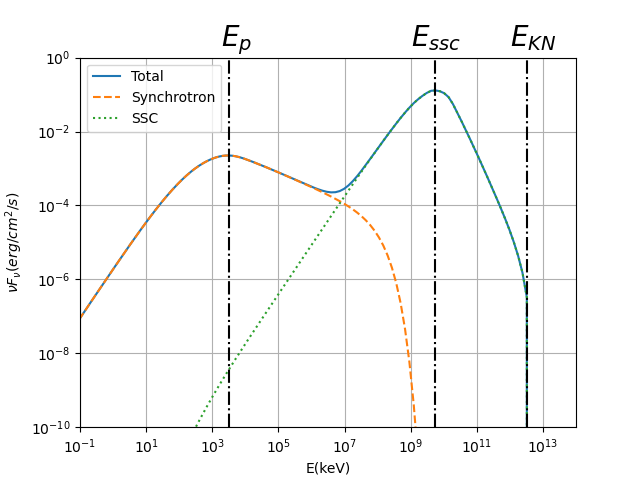}}
    \subfigure[Different shapes \label{fig:patterns}]{
        \includegraphics[width=0.45\textwidth]{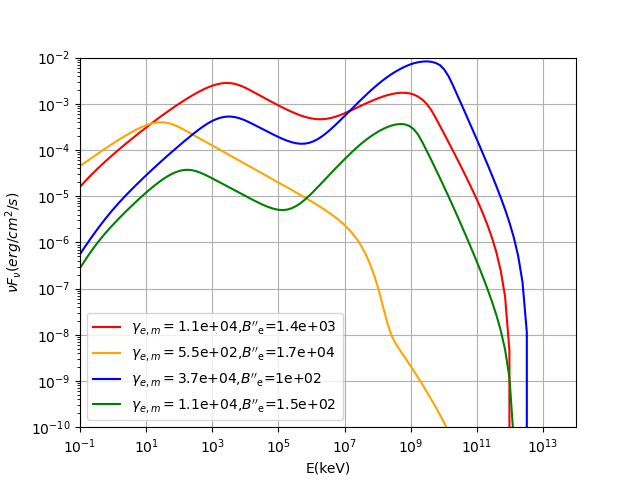}}
    \caption{(a)The solid line represents the total spectrum, the dashed line and the dotted line represents the synchrotron component and the SSC component respectively. Parameters set up: $M_\text{bulk}= 10^{34}\rm g$, $\Gamma_0=84$, $\sigma_0=60$, $f_\text{p}=0.04$, $k=5\times10^{-11}$, $f_\text{e}=0.9$, $L'_0=6\times10^{10} \text{cm}$, $v'_\text{in}=1.5\times10^{9} \text{cm/s}$, $R_0=10^{14} \text{cm}$, $b=30$, $l=0.2$, each reconnection within the event has a distinct position and orientation characterized by $\theta \in (0, 2/\Gamma_0)$ and $\phi \in (0, \pi/2)$, the observation time is settled at the peak luminosity time; (b)Spectrums with different shapes.}
\end{figure}

\begin{figure}
    \centering
    \subfigure[$\gamma_{\rm e,m} = 10^3$]{
        \includegraphics[width=0.3\textwidth]{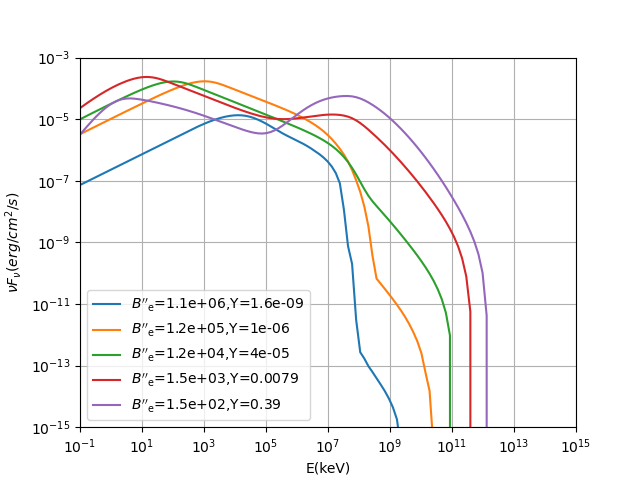}}
    \subfigure[$\gamma_{\rm e,m} = 10^4$]{
        \includegraphics[width=0.3\textwidth]{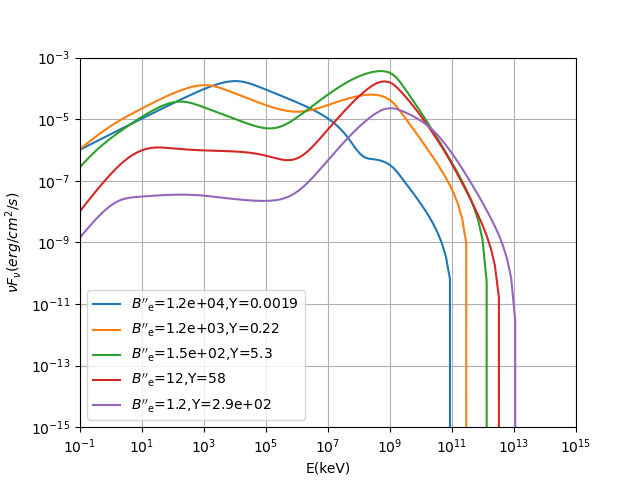}}
    \subfigure[$\gamma_{\rm e,m} = 10^5$]{
        \includegraphics[width=0.3\textwidth]{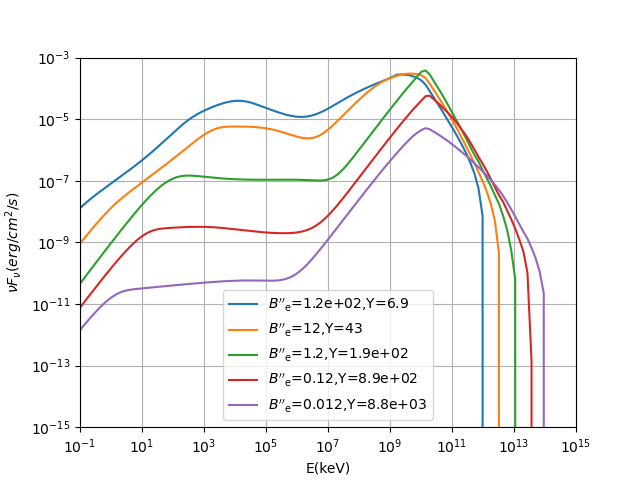}}
    \caption{Spectrums with different $\gamma_{\rm e,m}$ and ${B''}_{\rm e}$. \label{fig:SSC cooling rules}}
\end{figure}


To investigate the spectral characteristics, we perform a series of comprehensive simulations within the framework of the ICMART model.
Simulation results demonstrate that the spectral shape exhibits a variety of patterns under different parameters settings.
We summarize the key findings as below.
We begin by presenting a representative spectral shape, as shown in \cref{fig:sample}.
As proved in \citet{Shao2022}, the synchrotron spectrum in the ICMART model is primarily determined by the magnetic reconnections oriented toward the observer.
This conclusion also extends to the SSC emission, allowing the overall spectrum to be decomposed into two distinct components: a synchrotron component and an SSC component.
The synchrotron component peaks at $E_{\rm p}$ and is characterized by a low-energy spectral index $\alpha$ ranging from $-1$ to $-1.5$.
The SSC component peaks at $E_{\rm ssc}$.
At the high-energy end, the SSC emission is suppressed by the KN effect above $E_{\rm KN}$, leading to an abrupt exponential cutoff in the strong KN regime or a more gradual decline in the weak KN regime.
The flux ratio between these two components is denoted as the the Y parameter.
Although the two components generally produce two distinct peaks, the SSC peak can be suppressed when the Y parameter is small, rendering only the synchrotron peak observationally prominent.
Depending on the parameter choices, the combined spectrum can exhibit a variety of shapes, as shown in \cref{fig:patterns}.

Since the Y parameter is governed by $\gamma_{\rm e,m}$ and ${B''}_{\rm e}$, we vary $E_{\rm p}$ across its observed range (1--$10^4$keV) and systematically examine the dependence of the Y parameter on $\gamma_{\rm e,m}$ and ${B''}_{\rm e}$ in \cref{fig:SSC cooling rules}.
We find that the Y parameter always increases as ${B''}_{\rm e}$ decrease.
This occurs because a lower ${B''}_{\rm e}$ weakens synchrotron cooling, resulting in an electron population with higher energies, which in turn enhances the SSC process.
Moreover, for the same fractional reduction in ${B''}_{\rm e}$, a system with a larger initial ${B''}_{\rm e}$ produces a greater rise in the Y parameter than the smaller ${B''}_{\rm e}$.
In contrast, the Y parameter also increases with $\gamma_{\rm e,m}$, but its increase shows minimal dependence on the initial value of $\gamma_{\rm e,m}$.

In the ICMART model, the event enters into a deep fast cooling regime when the critical cooling electron Lorentz factor,
\begin{equation}
    \gamma_{\rm e,c}=\frac{6\pi m_e c}{\sigma_T(1+Y){B''}_{\rm e}^2\Delta t''}\ll\gamma_{\rm e,m}
\end{equation}
where $\Delta t''$ is the duration time of the magnetic reconnection.
This regime typically produces a synchrotron spectrum with $\alpha\approx-1.5$.
Observations however, show that some GRBs exhibit bigger $\alpha$, sometimes approaching $-1$.
This deviation from the canonical fast-cooling slope can be explained by the synchrotron radiation in a weak and decaying magnetic field \citep{Uhm2013,Shao2022}.
In such cases, high $\gamma_{\rm e,m}$ will be required to keep $E_{\rm p}$ within the observed range since $E_{\rm p}\propto\gamma_\text{e,m}^2{B''}_{\rm e}$.
To maintain $\gamma_{\rm e,c} \lessapprox \gamma_{\rm e,m}$ when SSC cooling is considered, ${B''}_{\rm e}$ must be sufficiently weak to compensate for the cooling enhancement brought by a large $Y$ parameter, which itself grows with weaker ${B''}_{\rm e}$ and larger $\gamma_{\rm e,m}$.
Consequently, an $\alpha$ higher than $-1.5$ implies a Y parameter larger than $1$ and a strong KN suppression due to the elevated $\gamma_{\rm e,m}$ \citep{Nakar2009}, as illustrated by the blue, orange and green lines in the right-most panel of \cref{fig:SSC cooling rules}.
We further note that for ${B''}_{\rm e} < 10\text{G}$, the resulting synchrotron spectrum begins to deviate markedly from the typical Band shape, as indicated by the purple line in the middle panel and the red and purple lines in the right-most panel of \cref{fig:SSC cooling rules}.
This is because an extremely weak magnetic field can lead to $\gamma_{\rm e,c} > \gamma_{\rm e,m}$, pushing the system into the slow cooling regime.

Within the ICMART model, $\gamma_{\rm e,m}$ and ${B''}_{\rm e}$ are related to $k$, $f_{\rm e}$ and $\sigma_0$ as follows \citep{Shao2025}:
\begin{align}
    {B''}_{\rm e} &\propto \sqrt{k}\sigma_0,\\
    \gamma_\text{e,m}&\propto\sqrt{\sigma_0}/f_{\rm e}.
    \label{eq:parameters connection}
\end{align}
Consequently, these three parameters of the ICMART model jointly determine the resulting spectral properties:
\begin{itemize}
    \item An increase in $k$ enhances ${B''}_{\rm e}$, driving a lower Y parameter;
    \item An increase in $f_{\rm e}$ reduces $\gamma_\text{e,m}$, thereby the Y parameter;
    \item An increase in $\sigma_0$ increases both ${B''}_{\rm e}$ and $\gamma_\text{e,m}$, leading to higher $E_{\rm p}$.
\end{itemize}

\subsection{Simulation of GRB 221009A}
\begin{table}
    \begin{tabular}{c c c c|c c|c}
        \toprule
        $\sigma_0$ & $f_\text{p}$ & $k$ & $f_\text{e}$ & $\gamma_\text{e,m}$ & $B''_{\rm e}(\rm Gauss)$ & Y \\
        \midrule
        4 & 0.53 & $2\times10^{-3}$ & 0.2 & $2.7\times10^3$ & $1.1\times10^4$ & $0.0055$ \\
        10 & 0.22 & $3\times10^{-4}$ & 0.45 & $2.5\times10^3$ & $1.5\times10^4$ & $0.0049$ \\
        20 & 0.11 & $7\times10^{-5}$ & 1 & $2.1\times10^3$ & $5.8\times10^3$ & $0.0058$\\
        \midrule
        30 & 0.07 & $1\times10^{-5}$ & 1 & $2.1\times10^3$ & $2.7\times10^3$ & $0.015$ \\
        40 & 0.053 & $5\times10^{-6}$ & 1 & $2.5\times10^3$ & $1.7\times10^3$ & $0.045$ \\
        60 & 0.035 & $3\times10^{-7}$ & 1 & $3.1\times10^3$ & $8.4\times10^2$ & $0.41$ \\
        \bottomrule
    \end{tabular}
    \caption{Parameters for \cref{fig:ICMART_spectrum_search_factor}, $M_\text{bulk}=10^{30} \rm g$ and $\Gamma_0=84$.}
    \label{table:final set}
\end{table}

\begin{figure}
    \centering
    \includegraphics[width=\textwidth]{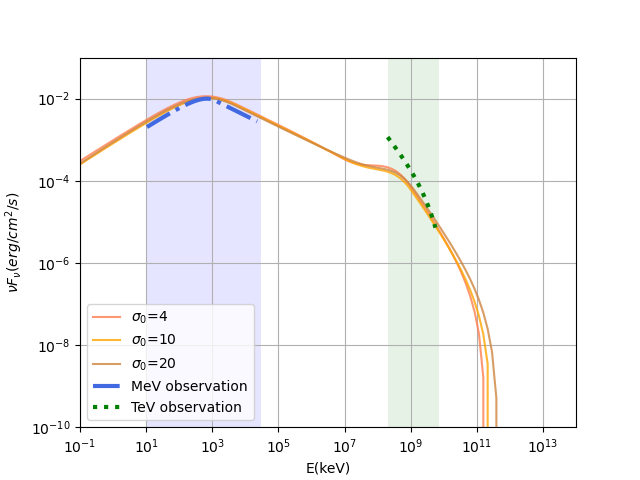}
    \caption{Solid lines with different shades of browns represent different $\sigma_0$ listed in the label. The blue dashed-dotted lines are the fitting functions of the observed MeV spectrums and the green dotted line is the fitting function of the observed TeV spectrums in $T_0+[240, 250] \rm s$.}
    \label{fig:ICMART_spectrum_search_factor}
\end{figure}

GRB 221009A is the brightest GRB ever recorded, with a peak luminosity reaching up to $1.7 \times 10^{54} \text{erg/s}$ \citep{HXMT}.
In this work, we focus on the observations of GRB 221009A within $T_0+[230, 250]\rm s$ ($T_0$ marks the Gamma-ray Burst Monitor (GBM) trigger time).
This interval comprises two time windows ($T_0+[230, 240]\rm s$ and $T_0+[240, 250]\rm s$) during the main emission phase that exhibit stable MeV emission and have simultaneous TeV coverage \citep{HXMT, LHAASO}.
During this time interval, a transient, narrow emission observed in the MeV range is believed to be generated by electron-positron pair annihilation, indicating an initial Lorentz factor of the bulk $\Gamma_0 = 84$ \citep{Ravasio2023,Zhang2024}.
The final Lorentz factor of the bulk acceleration can be calculated as \citep{Zhang2018book,LHAASO}:
\begin{align}
    \Gamma_\text{f} &= 0.9^{3/8}(\frac{3E_\text{k}(1+z)^3}{\frac43\times2\pi nm_pc^5t_\text{p}^3})^{1/8} = 578 E_\text{k,55}^{1/8}n_0^{-1/8}(\frac{t_\text{p}}{18\text{s}})^{-3/8},
\end{align}
where $E_\text{k}$ is the isotropic kinetic energy of the ejecta, $n$ is the number density of the circum-burst medium, $m_p$ is the mass of the proton and $t_\text{p}\sim18\text{s}$ is the peak time of the light curve.
The value of $\Gamma_\text{f}$ imposes a constraint on the amount of magnetic reconnections, which subsequently limits the possible range of $f_\text{p}$.
Combined with the measured luminosity, this allows the mass of the bulk to be estimated as $10^{30}\rm g$.

In $T_0+[240, 250]\rm s$, MeV observations from Insight-HXMT and GECAM-C reveal that $E_{\rm p}=683 \text{keV}$ and the low-energy spectrum index $\alpha=-1.5$ \citep{HXMT}.
The flux ratio between TeV and MeV is measured as $2\times10^{-5}$ \citep{LHAASO}.
Considering the $\gamma\gamma$ absorption and the suppression factor due to SSC peak deviates from the LHAASO energy range, the ratio of SSC to synchrotron luminosity ratio is estimated to be less than $6\times10^{-3}$ \citep{Dai2023}.
We explore a parameter space with $\sigma_0$ ranging from $4$ to $60$ while systematically adjusting other model parameters to achieve optimal spectral fitting to the observational data.
As demonstrated in subsection \ref{subsec:spectral shape}, an increase in $f_{\rm e}$ leads to a reduction in the Y parameter. Consequently, for a given $\sigma_0$, the Y parameter reaches its minimum when $f_{\rm e} = 1$.
Simulation results indicate that scenarios with $\sigma_0 > 20$ have  minimum Y parameters larger than the observed value of $Y = 6\times10^{-3}$, as shown in the lower part of \cref{table:final set}.
In contrast, spectrums with $\sigma_0 \leq 20$ exhibit excellent agreement with MeV observations while not exceed TeV measurement, as visually demonstrated in the spectral comparison presented in \cref{fig:ICMART_spectrum_search_factor} and the corresponding parameter sets are listed in the upper part of \cref{table:final set}.
The discrepancy between the simulated results and the observation in TeV band should be attributed to contamination from external shock radiation.

On the other hand, MeV observations in $T_0+[230, 240]\rm s$ report $E_{\rm p}=1200 \text{keV}$ and $\alpha=-1$ \citep{HXMT}.
According to our simulation results, such a high $\alpha$ indicates a low ${B''}_{\rm e}$, leading to a large Y parameter at least three orders of magnitude higher than the observed level.
Therefore, based on this observation, we are unable to find suitable parameters within the ICMART model framework to explain the data.

\section{Conclusion and Discussion}
\label{sec:conclusion}
In this study, we numerically compute the synchrotron self-Compton (SSC) spectrum within the ICMART model using the dedicated code \ICMARTpy, and systematically examine how key physical parameters influence the resulting spectral energy distribution.

The overall spectral shape is fundamentally governed by three independent parameters in the ICMART model, i.e. $k$, $f_{\rm e}$ and $\sigma_0$, through their relationships with the minimum electron Lorentz factor $\gamma_\text{e,m}$, and the comoving magnetic field strength ${B''}_{\rm e}$. These quantities, in turn, determine the relative strength and morphology of the synchrotron and SSC spectral components.

We find that the broadband spectrum typically exhibits two distinct peaks. However, both the relative prominence and the relative position of these peaks vary significantly with physical conditions:
\begin{itemize}
\item When the Y parameter is typically small due to low $\gamma_\text{e,m}$, the SSC component is suppressed and often obscured by the dominant synchrotron peak.
\item At high energies, the SSC spectrum can exhibit a sharp cutoff above $E_{\rm KN}$ due to Klein-Nishina suppression, particularly when $\gamma_\text{e,m}$ is large.
\end{itemize}
Our analysis further reveals that the Y parameter increases with a decreasing ${B''}_{\rm e}$ and an increasing $\gamma_\text{e,m}$.
This dependency gives rise to an observational link between synchrotron and SSC spectral properties:
a synchrotron spectrum with $\alpha > -1.5$ suggests a weak ${B''}_{\rm e}$ and a large $\gamma_\text{e,m}$, indicating a large Y parameter and a strong KN regime.

Employing the ICMART model, we perform detailed simulations of the broadband spectral energy distribution of GRB 221009A with the aim of identifying physically plausible parameters that can explain its observed MeV--TeV emission.
A key finding from our modeling is the relationship between the Y parameter and the magnetization parameter $\sigma_0$.
To maintain consistency with the observed peak energy, the Y parameter must scale as (see Appendix for details)
\begin{equation}
    Y \propto \sigma_0^{1+\frac{\eta}{2}},
\end{equation}
indicating a positive correlation in both weak and strong Klein-Nishina regimes.
This trend contrasts with the anti-correlation predicted by the internal shock model, a discrepancy arising from differences in the expressions for $\epsilon_e/\epsilon_B$ between the two models.

By comparing our synthetic spectra with multi-wavelength observations during $T_0+[240, 250]\rm s$, we find that the Y parameter exceeds the observationally allowed values when
$\sigma_0>20$.
We therefore conclude that $\sigma_0\leq20$ is required to remain consistent with the empirical constraints.
We note that the observed flux is systematically higher than our simulated values, which may be due to contamination from afterglow emission.
This behavior of the $Y-\sigma_0$ relation underscores a key signature of magnetic reconnection-dominated models such as ICMART.

During the interval $T_0+[230, 240]\rm s$, we find that no parameter set within the ICMART model can adequately reproduce the observed spectral properties.
In particular, the predicted Y parameter substantially exceeds the observationally inferred value by approximately three orders of magnitude.

While this discrepancy might reflect a challenge to the ICMART model in its current form, it is important to note that other plausible explanations exist.
For instance, $\alpha$ may appear elevated due to the temporal averaging of the spectrum over approximately $10$ seconds \citep{LHAASO}.
Alternatively, the emission during this interval may include a contribution from a mechanism not considered in our current model, such as a thermal component.
If present, the thermal component could provide substantial flux in the low-energy band, thereby raising the observed $\alpha$.
Such a component would likely originate from an earlier evolutionary phase, which explains why a single-component non-thermal model successfully fits the spectrum in the subsequent interval.
We note, however, that any such thermal component must remain spectrally subdominant in our analysis, as the current broadband data lack the clear diagnostic features required to robustly decompose and identify it as a separate spectral entity.
More generally, our result implies that any GRB exhibiting both a high observed $\alpha$ and a low measured $Y$ parameter likely requires an additional emission component beyond pure synchrotron radiation to account for the enhanced low-energy flux.


Overall, our results demonstrate a systematic correlation between MeV and TeV spectral components in GRBs, highlighting the crucial role of broadband observation in deciphering radiation mechanisms and energy dissipation processes during the prompt phase.
Future coordinated observations between TeV instruments such as LHAASO, CTA (Cherenkov Telescope Array) \citep{CTA}, and MAGIC (Major Atmospheric Gamma Imaging Cherenkov Telescope) \citep{MAGIC}, and gamma-ray satellites including Fermi (Fermi Gamma-ray Space Telescope) \citep{Fermi}, Swift, and SVOM (Space Variable Objects Monitor) \citep{SVOM} will provide unprecedented multi-wavelength coverage.
Such datasets will be essential to rigorously test and refine emission models like ICMART, ultimately advancing our understanding of the physical processes powering gamma-ray bursts.

\begin{acknowledgments}
    We sincerely thank Prof. Bing Zhang for his continuous engagement in scientific discussions and his constructive suggestions throughout the development of this study, which significantly contributed to this paper.
    We are also grateful to Prof. Giancarlo Ghirlanda for his insightful comments, which greatly improved our methodological approach.
    Finally, we acknowledge the anonymous reviewers for their valuable feedback.
    This work is supported by the National Natural Science Foundation of China (Projects 12373040,12021003) and the Fundamental Research Funds for the Central Universities.
\end{acknowledgments}

\appendix
\section{Estimation of SSC Spectral Parameters}
\label{appendix:estimation}
The SSC component is suppresses by the KN effect, causing both $E_{\rm ssc}$ and $E_{\rm KN}$ deviate from Thomson prediction.
In this paper, we adopt the analytical estimate for $E_{\rm ssc}$ from \citet{Nakar2009}:
\begin{equation}
    E_{\rm ssc}\approx2\Gamma_{\rm f}\gamma\gamma_\text{e,m}m_{\rm e}c^2.
\end{equation}
$E_{\rm KN}$ is determined by the maximum energy of the injected electron population:
\begin{equation}
    E_{\rm KN} = \Gamma_{\rm f}\gamma_\text{e,M}m_{\rm e}c^2 \propto {B''}_{\rm e}^{-\frac12}.
\end{equation}

In addition to the characteristic spectral energies, we introduce the Y parameter to quantify the relative strength of the SSC to synchrotron components.
\citet{Nakar2009} has estimated the Y parameter as
\begin{align}
    Y \approx \sqrt{\frac{\epsilon_e}{\epsilon_B}}
    \begin{cases}
        1 & \frac{\gamma_\text{e,m}}{\hat{\gamma_\text{e,m}}}<p-2,\\
        \sqrt{p-2}(\frac{\gamma_\text{e,m}}{\hat{\gamma_\text{e,m}}})^{-1/2} & \frac{\gamma_\text{e,m}}{\hat{\gamma_\text{e,m}}}>p-2,
    \end{cases}
    \label{eq:Y in Nakar}
\end{align}
where $\hat{\gamma_\text{e}} = \Gamma\gamma m_{\rm e}c^2/(h\nu_{\rm syn}(\gamma_\text{e}))$ and $\nu_{\rm syn}(\gamma_\text{e})$ is the characteristic frequency of $\gamma_\text{e}$; $\epsilon_e$ and $\epsilon_B$ are the electron energy and the radiation magnetic field energy separately.
Within the ICMART regime, their ratio can be estimated as
\begin{equation}
    \frac{\epsilon_e}{\epsilon_B} = \frac{4\pi Q_0m_ec^2\gamma_\text{e,m}^2}{(p-2){B''}_{\rm e}^2{L''}^3},
\end{equation}
where $L''$ is the size of the reconnection area in the mini-jet comoving frame.
Therefore, \cref{eq:Y in Nakar} can be written as
\begin{equation}
    Y \propto {B''}_{\rm e}^{-1}\gamma_\text{e,m}^{\eta} \label{eq:Y in ICMART},
\end{equation}
where $\eta$ is $1(-1/2)$ for the weak(strong) KN regime.
However it should be cautioned that the analytical estimation described above may deviate from numerical results.
For instance, the approximation of $E_{\rm KN}$ can introduce an uncertainty of approximately one order of magnitude compared to numerical simulations.
Therefore, analytical results derived from such approximations should be regarded primarily as a reference for topological or qualitative analysis, rather than for precise quantitative interpretation.

\bibliography{paper.bib}

\end{document}